\documentclass{article}
\usepackage{amsmath}
\usepackage{amssymb}
\usepackage{makeidx}
\usepackage{latexsym}
\usepackage[dvips]{graphicx}
\usepackage{textcomp}

\begin{document}

\title{{\huge Geometrical Lorentz Violation}\\
and \\
Quantum Mechanical Physics}
\author{{\ Roberto Mignani}$^{2,3,4}${\ , Andrea Petrucci}$^{2,5,*}${\ and Fabio Cardone}$^{1,2}$ \\
\\
$^{1}$Istituto per lo Studio dei Materiali Nanostrutturati (ISMN -- CNR)\\
Via dei Taurini - 00185 Roma, Italy\\
$^{2}$GNFM, Istituto Nazionale di Alta Matematica "F.Severi"\\
\ Citt\`{a} Universitaria, P.le A.Moro 2 - 00185 Roma, Italy\\
$^{3}$Dipartimento di Matematica e Fisica \\
Sezione di Fisica \\
Universit\`{a} degli Studi \textquotedblright Roma Tre\textquotedblright \\
\ Via della Vasca Navale, 84 - 00146 Roma, Italy\\
$^{4}$\,I.N.F.N. - Sezione di Roma Tre \\
$^5$ ENEA-Casaccia, via Anguillarese 301 - 00123 Roma, Italy\\
$^{*}$corresponding author - petrucciandr@gmail.com +393392573280\\
\\
}
\maketitle
\date{}

\begin{abstract}
On the basis of the results of some experiments dealing with the violation
of Local Lorentz Invariance (LLI) and on the formalism of the Deformed
Special Relativity (DSR), we examine the connections between the local
geometrical structure of space-time and the foundation of Quantum Mechanics.
We show that Quantum Mechanics, beside being an axiomatic theory, can be
considered also a deductive physical theory, deducted from the primary
physical principle of Relativistic Correlation. This principle is synonym of
LLI and of a rigid and flat minkowskian space-time. The results of the
experiments mentioned above show the breakdown of LLI and hence the
violation of the principle of Relativistic Correlation. The formalism of DSR
allows to highlight the deep meaning of LLI breakdown in terms of the \emph{%
geometrical structure of local space-time} which, far from being rigid and
flat, is deformed by the energy of the physical phenomena that take place
and in this sense it has an active part in the dynamics of the whole
physical process. This perspective has a far reaching physical meaning that
extends its consequences to the foundations of Quantum Mechanics according
to the interpretation of Copenhagen. It provides a 'real' explanation and
description of quantum phenomena enriching, by the concept of deformed
space-time, the realistic interpretation in terms of pilot wave and hence it
uncovers the reality hidden below the probabilistic interpretation and
dualistic nature of quantum objects.
\end{abstract}

\section{The principle of relativistic correlation}

The study of the laws of Nature has always found in the cause-effect
relationship (causality), between two events, a very powerful method of
investigation. Causality stands on the principle of relativistic correlation
which establishes the temporal order of events. Before the experimental
disclosure of the deep link between electricity and magnetism and its
formalisation in terms of Maxwell's equations and before the understanding
that these equations described the interactions among elementary particles
(that had just been discovered), physics relied on the not-physical concept
of "action at a distance". Maxwell's equations formalised the concept of
field as conveyor of the electromagnetic interaction among bodies and being
its propagation speed the speed of light, it became possible to establish
causality at a finite speed. However, a further step forward had to be moved
before defining the principle of relativistic correlation. This step had to
be moved through the whole process that brought from Galilean Relativity to
Special Relativity or, rather, to the Einsteinian Relativity. The postulate
of relativity and the postulate of a universal limiting speed~\footnote{%
Of the two postulates, the latter has been the crucial one for the
development of Quantum Mechanics and for the following development of whole
Physics, both theoretical and experimental until nowadays. It is also called
"postulate of the constancy of the speed of light" but "because special
relativity applies to everything not just light, it is desirable to express
it in terms that convey its generality"~\cite{Jackson}. This is how
J.D.Jackson expresses the importance of this postulate, but it reflects also
the point of view of a young Einstein who, moved by idealism, aimed at
making universal experimental evidences, that despite their strong
accountability, were strictly true only for electromagnetism.}, by Einstein,
contain, as direct consequences, both Lorentz transformations and the flat
and rigid Minkowski space-time. The former implies that the speed of light
is the limiting speed for physical phenomena, the latter implies that c is
the maximal causal speed. From these two points of view, it turns out that
the causal velocity is not just \emph{finite} (i.e. not infinite), but also
\emph{limited} (i.e. it cannot be bigger than a certain value), \emph{unique}
(i.e. valid for any interaction and energy independent), \emph{constant}
(i.e. time independent in an inertial reference system), \emph{invariant}
and coinciding with the numerical value of the light speed in vacuum and in
this last sense maximal.

\section{Quantum Mechanics as deductive theory}

Quantum Physics relies on the proposal of three new phenomenogical models
based on revolutionary concepts: quantization of energy exchanges between
radiation and matter, by Max Planck; quantization of the electromagnetic
radiation and indivisibility of space and time by Albert Einstein;
quantization of the energy of atomic electrons by Niels Bohr. From these
assumptions, as stated by M. Jammer in~\cite{concepdevQM}, in the initial
years, Quantum Physics had developed as a "deplorable patchwork of
hypotheses, principles, theorems and computational rules" that allowed to
match the predictions, obtained through classical methods, with experimental
data. Several theoretical works were developed in order to systematize this
empirical methods. In 1924 de Broglie showed the wave nature of material
particles, in 1925 Heisenberg developed together with Born and Jordan a
formalism to face quantum problems known as Matrix Mechanics and a few
months later Dirac came out with his formalism called quantum algebra which
produces the same results as Heisenberg's. Eventually, in 1926, it appeared
the formalism developed by Schr\"{o}dinger which initiated the version of
Quantum Physics known as Wave Mechanics. The conference held in Como in
September 1927 may certainly be considered as the conclusion of the first
period of theoretical construction of the Quantum Theory and as the birth of
Quantum Mechanics as an axiomatic theory\footnote{%
This statement means that Quantum Mechanics is not based on any prime
physical principle, but only on a coherent set of postulates from which
operative conditions can be extrapolated and predictions matching
experiments can be obtained. These are the postulates about the wave
functions, the observables, the Hermitian operators, the probability
interpretation, the complete set of independent eigenfunctions, the
expectation values, the time evolution of the wave function.}, according to
the "Copenhagen Interpretation". From an historical perspective the process
of development of Quantum Mechanics went on in the following years when the
first steps towards a quantum electrodynamics began to be moved. In 1928
Dirac published the first relativistic wave equation for the electron and in
the same year Jordan and Wigner wrote the relativistically invariant
commutation relations for the electromagnetic field. This is the most
important result of this formalism, from a theoretical point of view,
because they are considered the quantization relations and from an
experimental and epistemological point of view because from them the
indeterminacy relations can be inferred. Despite the important results, the
formalism of quantum electrodynamics presented several considerable problems
like divergences and negative energy states. These difficulties produced
several doubts about the soundness of the formalism. Because of these
problems, several physicists, who had contributed to the foundation of the
formalism, like Heisenberg, Bohr and Rosenfeld, wanted, some years later, to
deeply experimentally analyse the predictions of the theory in order to
prove its soundness. Above all Bohr and Rosenfeld wanted to test the
predictions of the Quantum Mechanics with infinite degrees of freedom, where
Spacial Relativity has to be explicitly used. They focused their attention
on the analysis of complementarity and on the commutation relations. From
the commutation relations for the electromagnetic field, in which the field
components refer to precise spacial points and hence do not possess a clear
and straight physical meaning, they designed some Gedankenexperimente, that,
despite this, were perfectly realisable. In these experiments, they
considered only the mean values of the fields over finite regions of
space-time. From these experiments, they obtained the same indeterminacy
relations that can be obtained from the commutation relations of the
formalism, and hence provided a physical foundation to the theory. This
process of deeper reanalysis of the experimental foundations of the
formalism of Quantum Electrodynamics, that was carried out by the founding
fathers of this theory, is a precious work as it discloses new aspects of
the existing relation between the relativistic and the quantum-mechanical
descriptions of physical phenomena and moreover it sheds a possibly
clarifying light on the heart of Quantum Mechanics that makes clear its
limits of validity. By considering the book by Heisenberg~\cite{physprincQM}
and some papers by Bohr and Rosenfeld~\cite{kalchar} in which they describe
their process of reanalysis, mentioned above, it clearly emerges that all
their Gedankenexperimente and hence the indeterminacy relations they get to%
\footnote{%
We remind again that these indeterminacy relations are equal to those
obtained from the commutation relations of the mathematical formalism and
that this equality established the physical foundations of this formalism.},
have a common physical foundation: the primary principle of relativistic
correlation\footnote{%
From the previous paragraph, it is clear that this principle is synonym of
Special Relativity and of flat and rigid Minkowskian space-time}. Without
giving too many details of the Gedankenexperimente that these physicists
imagine in order to determine the mean values of the components of the
electromagnetic fields and their indeterminacy relations, we will only
sketch their main ideas and show that they are clearly always compatible
with the relativistic correlation. Heisenberg is convinced that the
classical concepts of particle (position and velocity) and wave need to be
used in the description of a quantum experiment too and he shows by
Gedankenexperimente that in the quantum world these two concepts are
inadequate to achieve a cause-effect relation and that their concomitant
application to the same microscopical experiment brings about the
indeterminacy relations. It is evident that these concepts (corpuscle and
wave) were stably related, since their birth, to an isotropic and
homogeneous space (the Euclid space) and then they were generalized, after
the advent of Special Relativity, to continue to be valid in the flat and
rigid minkowskian space-time. In his Gedankenexperimente, he chooses cubic
volumes and right angles, where to evaluate the energy and intensity of the
electromagnetic field which, of course, is expressed by the Maxwell's
equations or expressions derived from them (Poynting vector, energy density,
Li\'{e}nard--Wiechert potentials or, equivalently, the fields obtained by a
Lorentz boost applied to a Coulomb electric field). It goes without saying
that all of these ideas are deeply rooted in the relativistic correlation.
Similar kinds of considerations are used by Bohr. He is convinced that the
physically meaningful statements of the theory are those regarding the
average values of the components of the electromagnetic field and, hence,
that the mathematical formalism is an idealization that acquires physical
meaning when integrated over space-time. With regards to the
Gedankenexperimente, he states several times that they have to be treated
from a classical point of view considering only classical concepts. He
considers extended bodies, not point charges, in order to reduce to zero the
radiation emitted by them (radiation reaction) during the measurement of
their momentum and he states that due to the finite (not infinite) speed of
light the body cannot be considered rigid, but it must be made of many parts
interconnected with each other by springs. He bases his considerations and
measurements on the classical formalism of Maxwell's equations (flat and
rigid minkowskian space-time) and on the classical concepts. Then, he
considers the electromagnetic field as a quantum object, i.e. made up of
photons and hence, he considers its corpuscular nature. Photons are emitted
according to the Poisson distribution which foresees their probability of
being emitted. He shows that the mean value of this distribution coincides
with the field measured in classical conditions in the Gedankenexperiment
and that the distribution possesses fluctuations which never cancel even
when the distribution is identically zero. Exactly as Heisenberg did, he
also shows that the results of the Gedankenexperimente are the indeterminacy
relations for the components of the field that are precisely compatible with
those obtained by integrating, over a minkowskian space-time, the
commutation relations of the theoretical formalism\footnote{%
As to these indeterminacy relations, it is interesting to know that, at the
beginning, Heisenberg called them uncertainty relation and only later Bohr
changed their name to indeterminacy. While the word uncertainty is a
probabilistic concept with no ontological meaning, the word indeterminacy is
an epistemological concept which states the impossibility of knowledge.}. It
is clear that both the experimental part and the integrating part have the
relativistic correlation as their groundwork. From the perspective gained
through the deep revisions made by Heisenberg and Bohr about the physical
foundations of Quantum Mechanics, it becomes possible to consider it, not
only an axiomatic theory, but also a deductive theory deducted from the
principle of relativistic correlation.

\section{Relativistic correlation, Local Lorentz Invariance and Space-Time
structure}

After showing the physical foundation of the formalism of Quantum Mechanics%
\footnote{%
The formalism of Quantum Mechanism (the commutation relations) which comes
from axiomatic assumptions, was proved to be physically sound because the
results obtained by it could be obtained as well by realisable experiments.}%
, its founders were no longer worried about the paradoxes that still
remained in it like the instantaneous collapse of the wave function, the
ontological meaning of indeterminacy, the wave-corpuscle dualism and the
contrast between reality and locality. On the contrary, some other
physicists, like Einstein, de Broglie, Schroedinger, Dirac, and some years
later Bohm continued to be deeply dissatisfied with a theory that, despite
its capacity to produce predictions in agreement with the results of the
experiments, was only capable to calculate the probability to obtain a
result and could not say anything about the description of the physical
phenomenon which ended up to be considered ontologically indeterminate until
the actual measurement. It will now be illustrated that, by looking at the
relativistic correlation (which we have seen to be at the basis of Quantum
Mechanics) from the point of view of the Local Lorentz Invariance (LLI) and
the local geometrical structure of space-time, new interesting and far
reaching physical perspectives on Quantum Mechanics will open up, which will
provide physical explanations to the paradoxes and a new and physically rich
insight of the wave-corpuscle dualism.

\subsection{LLI breakdown and Deformed Special Relativity}

As above said, the final step for the birth of the principle of relativistic
correlation was moved through the process that brought from the relativity
of Galilei to the Einsteinian Special Theory of Relativity. In other words,
it is not limiting at all to focus the attention on the two postulates of
Special Relativity rather than on the principle of relativistic correlation.
In particular, the attention has to be concentrated on LLI\footnote{%
It will be defined more precisely soon, but here it can be considered as the
synthesis of the two postulates mentioned above.} and its possible
breakdown. The concept of LLI comes from Einstein's relativity theories
which state that physical phenomena occur in a space-time whose structure is
globally curved (Riemannian) and locally flat (Minkowskian). The local
flatness of space-time means that the laws of physics can be locally written
in the language of Special Relativity (SR) and hence physical phenomena are
locally invariant under Lorentz transformations. The controversial point at
issue (from both the theoretical and the experimental side) is whether the
validity of local Lorentz invariance (LLI) is preserved at any length or
energy scale. It is worth mentioning that a great deal of attempts, both
theoretical and experimental, have been conducted so far from different
perspectives to predict LLI limits or measure them in order to find out some
possible signature of a new physics ~\cite%
{Will,Mattingly,Kostelecky,Camelia, Camelia1}. However,in order not to drift
away from the track that has been followed so far, they will not be
mentioned. The point of view, from which LLI breakdown is looked at and
dealt with in the theoretical and experimental research conducted so far, is
metrical. In other words, the questions to be addressed are 'how does the
Lorentz invariant Minkowskian Space-Time get deformed if LLI is broken?' and
'can this Space-Time deformation somehow affect the evolution of the
physical phenomena that take place in it?'. Two of the present authors,
Cardone and Mignani, started from the profound connection between the
breakdown of LLI and Space-Time geometry and parametrised the minkowskian
metric tensor by replacing the constant coefficients of \emph{%
diag(1,-1,-1,-1)} with coefficients depending on a phenomenological
parameter \emph{E} as in \emph{diag(b$_{0}$$^{2}$(E), -b$_{1}$$^{2}$(E), -b$%
_{2}$$^{2}$(E), -b$_{3}$$^{2}$(E))}. The parameter \emph{E} has the
dimension of energy and has to be interpreted as the energy exchanged during
the non-Lorentz invariant process\footnote{%
It is necessary to clarify the meaning of the words \textquotedblleft a
parameter with the dimension of energy which is considered to be the energy
exchanged during the non-Lorentz invariant process\textquotedblright . A
physical process which is not invariant under Lorentz symmetry (in the local
sense) takes places in or, more precisely, involves a locally non flat
spacetime. In order to detect the effects due to a non-flat spacetime
(either curved or deformed) it is necessary to perform two measurements and
then subtract their results. This is how Eddington operated when he measured
the deflection of the light of a distant star around the Sun by subtracting
two measured angles and this is how one operates with geodesic deviation to
determine the intrinsic curvature of a manifold. In our case, the parameter
\emph{E} parametrizes the local deformation of spacetime and, in this sense,
has to be understood as the difference of two energies measured under two
different conditions, as it happened for the angles in Eddington experiment.}%
. In order to determine the features of this parameter, the analytical forms
of the metric coefficients and other phenomenological details of this
theory, this formalism was used to analyse the experimental set-up and the
results of two experiments carried out in Cologne \cite{Nimtz} and Florence
\cite{Ranfagni} where LLI was broken in the sense that superluminal
propagation of electromagnetic waves was observed. From this analysis, it
was possible to extrapolate the energy function of the metric parameters ,
as it is shown in ~\cite{Energy and Geo, Deformed Space Time}. However, for
our purpose, that is to show the results of an experiment in which some
photons anomalously interact with other photons (not according QED), it is
sufficient to state that the analysis of the results of these two
experiments found out, first of all, an energy value of 4.5 $\mu $eV which
is the threshold value over which the local Space-Time becomes again
minkowskian\footnote{%
In all but very few experiments, LLI is valid and the laws of physics are
compliant with the language of special relativity. Hence, it goes without
saying that there must exist an energy threshold either very low or very
high (or maybe both) that acts as an upper bound or a lower bound
respectively for those energy ranges where LLI is broken. In our
phenomenological framework we search for a sufficiently low energy threshold
below which LLI is violated.} and second, the space extension and the angles
~\cite{Ranfagni} over which it is possible to make out the deformed
space-time effects of LLI breaking.

\subsection{The experiments}

Since we wanted to measure effects due to LLI breakdown, we designed the
experimental set-up according to the energy and space threshold mentioned
above. Moreover, since we were looking at LLI breakdown from the point of
view of the effects brought about by the locally deformed spacetime on the
propagation of photons, we had to search for these effects in a difference
of two measurements according to what has been explained in the previous
Section. Besides, since we wanted the results to shed new light on Quantum
Mechanics and, in particular, on the wave-corpuscle dualism, we decided to
design an experimental set-up where there would be the presence of the de
Broglie wave, i.e. a double slit like experiment. In the last decade we
carried out three experiments involving photon systems in the near infrared
range. Before moving on to the description we refer to Fig.1 where the
lay-out of the set-up is reported. The distances between sources and
detectors and in particular the distances L and s were fixed in the same
proportion as those of the Florence experiment \cite{Ranfagni} where LLI
breakdown showed up in the sense of superluminal propagation of
electromagnetic waves in the microwave range between two horn antennas. A
strong hypothesis was made here as to the independence of LLI breakdown from
the range of frequency (microwave to infrared).

\begin{figure}[tbp]
\begin{center}
\ \includegraphics[width=0.8\textwidth]{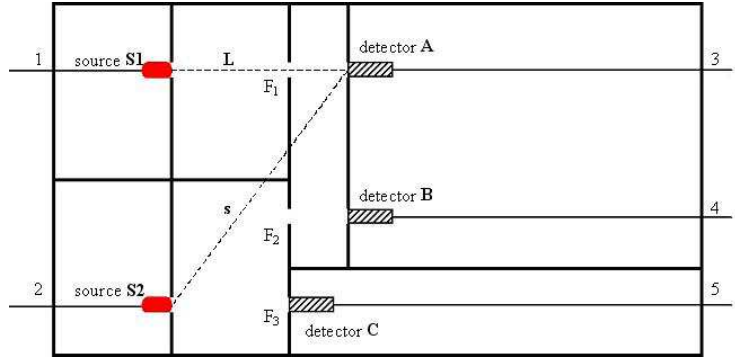}
\end{center}
\caption{Experimental set-up. The box contains two near infrared sources S1
and S2 and three detector A, B and C. The panels divide the room into
several vanes, some of them connected by apertures F1, F2 and F3.}
\label{exp_scheme}
\end{figure}

\subsection{Experimental set-up}

The apparatus employed in all experiments (schematically depicted in Fig.\ref%
{exp_scheme}) consisted of a Plexiglas box with wooden base and lid. The box
(thoroughly screened from those frequencies susceptible of affecting the
measurements) contained two identical infrared (IR) LEDs, as (incoherent)
sources of light, and three identical detectors (A, B, C). The two sources S$%
_{1}$, S$_{2}$ were placed in front of a screen with three circular
apertures F$_{1}$, F$_{2}$, F$_{3}$ on it. The apertures F$_{1}$ and F$_{3}$
were lined up with the two LEDs A and C respectively, so that each IR beam
propagated perpendicularly through each of them. The geometry of this
equipment and the absorbing material of the internal walls were designed so
that no photon could pass through aperture F$_{2}$ on the screen. The
wavelength of the two photon sources was $\lambda $ = 8.5$\cdot $10$^{-5}$
cm. The apertures were circular, with a diameter of 0.5 cm, much larger than
$\lambda $.We therefore worked in the absence of single-slit (Fresnel)
diffraction. However, the Fraunhofer diffraction was still present, and its
effects were taken into account in the background measurements. Detector C
was fixed in front of the source S$_{2}$; detectors A and B were placed on a
common vertical panel. Let us highlight the role played by the three
detectors. Detector C destroyed the eigenstates of the photons emitted by S$%
_{2}$. Detector B ensured that no photon passed through the aperture F$_{2}$%
. Finally, detector A measured the photon signal from the source S$_{1}$. In
summary, detectors B and C played a controlling role and ensured that no
spurious and instrumental effects could be mistaken for the anomalous
effect, which had to be revealed on detector A. The design of the box and
the measurement procedure were conceived so that detector A was not
influenced by the source S$_{2}$ according to the known and commonly
accepted laws of physics governing electromagnetic phenomena: classical
and/or quantum electrodynamics. In other words, with regards to detector A,
all went as if the source S$_{2}$ would not be there at all or would be kept
turned off all the time. In essence, the experiments just consisted in the
measurement of the signal of detector A (aligned with the source S$_{1}$) in
two different states of source lighting. Precisely, a single measurement on
detector A consisted of two steps: (1) Sampling of the signal on A with
source S1 switched on and source S$_{2}$ off; (2) sampling of the signal on
A with both sources S$_{1}$ and S$_{2}$ switched on. Analogous measurements
have been taken on detectors B and C. A possible non-zero difference $\Delta
$A = A (S$_{1}$ on S$_{2}$ off) - A(S$_{1}$ on S$_{2}$ on) in the signal
measured by A when source S$_{2}$ was off or on (and the signal in B was
strictly null) has to be considered as evidence for the searched anomalous
effect. Following all previous discussions, let us explicitly notice once
again that the geometry of the box was strongly critical in order to reveal
the anomalous photon behaviour.

\subsection{The results}

\begin{figure}[tbp]
\begin{centering}
\includegraphics{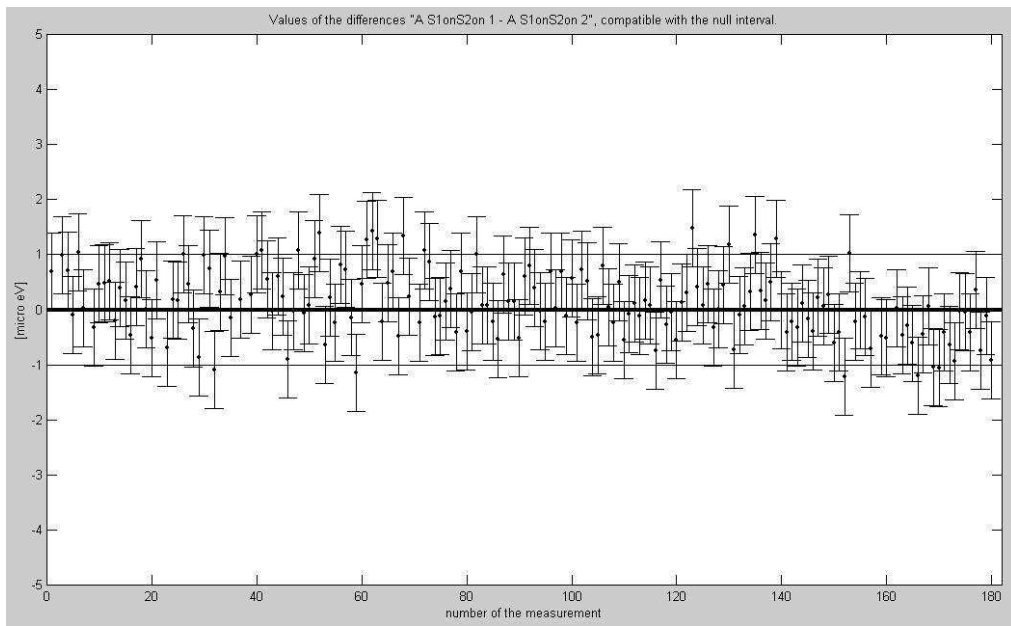}
\caption{\textit{Values of the differences} $\Delta A^{\prime}$
\textit{obtained by subtracting two sets of signal samples measured
on the detector A with both sources on. The differences are clearly
compatible with zero.}}\label{diff_compat_zero}
\par\end{centering}
\end{figure}

Three experiments were carried out by different sources, detectors, power
supplies and multimeters. The results of the first and second experiments
are reported in \cite{ombra1, ombra2, ombra2bis}. Only the results of the
third experiment is reported here. The third experiment was repeated several
times over a whole period of four months, in order to collect a fairly large
amount of samples and hence have a significant statistical reproducibility
of the results. Thanks to this large quantity of data, it was possible to
study the distribution of the differences of signals on detector A. We
considered two types of differences: the differences containing the
anomalous signal $\Delta $A = A (S$_{1}$ on S$_{2}$ off)-A (S$_{1}$ on S$_{2}
$ on) and the blank differences $\Delta $A$^{\prime }$ = A (S$_{1}$ on S$_{2}
$ on)-A (S$_{1}$ on S$_{2}$ on). In Fig.\ref{diff_compat_zero} we show that
the second type of differences are all compatible with zero (inside the zero
compatibility interval [-1; 1] $\mu $V).

\begin{figure}[tbp]
\begin{centering}
\includegraphics{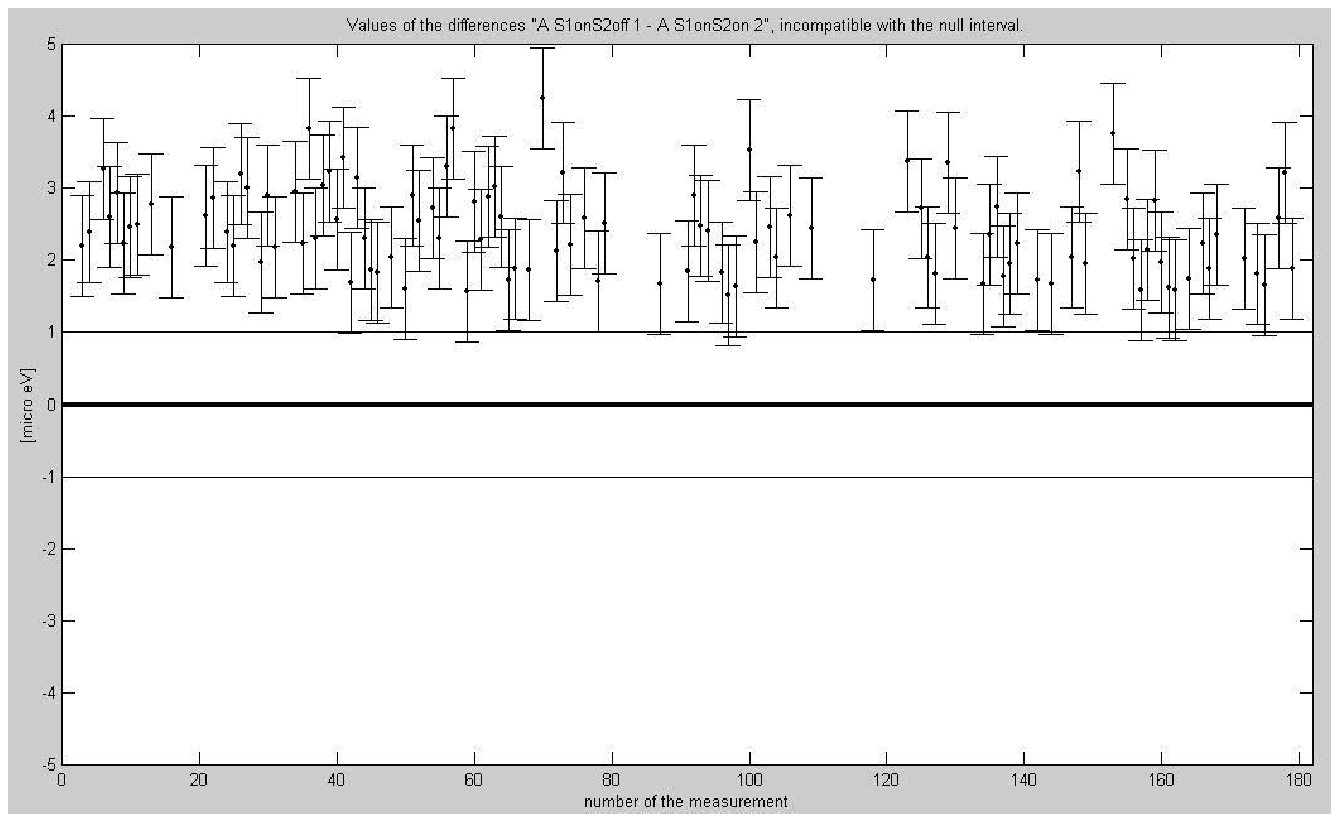}
\caption{\textit{Values of the differences} $\Delta A$ \textit{
obtained by subtracting two sets of signal samples measured on the
detector A with different lighting conditions of the sources:
S1onS2off and S1onS2on. The differences are clearly incompatible
with zero.}}\label{diff_no_compat_zero}
\par\end{centering}
\end{figure}

In Fig.\ref{diff_no_compat_zero} conversely, we show the differences of the
first type (A (S$_{1}$ on S$_{2}$ off)-A (S$_{1}$ on S$_{2}$ on)) not
compatible with zero after having discarded those difference whose error bar
was too much inside the interval [-1; 1] $\mu $V.

\begin{figure}[tbp]
\begin{centering}
\includegraphics{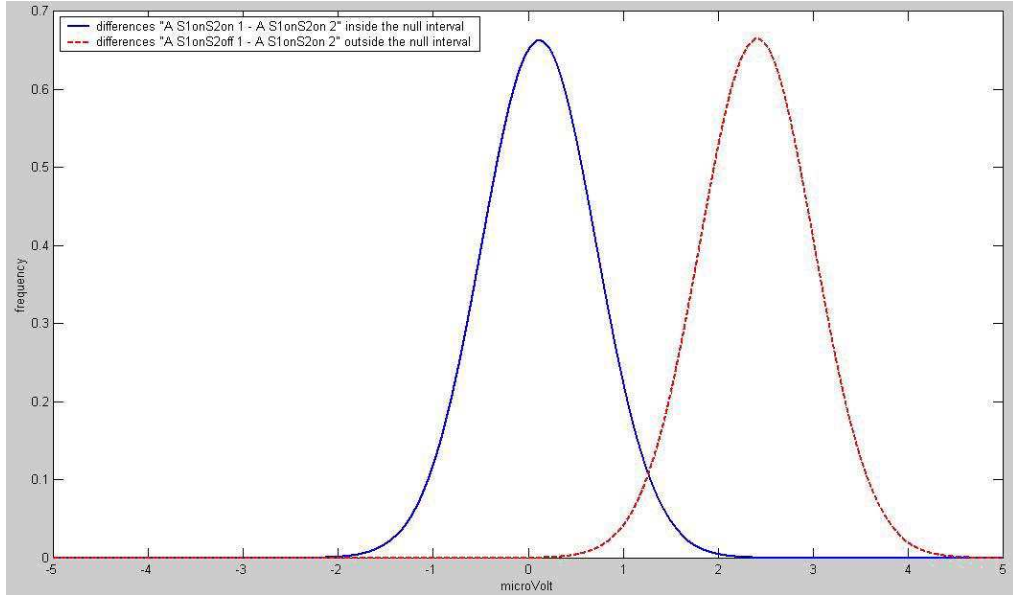}
\par
\caption{\textit{Gaussian curves (normal frequency vs. signal
difference in $\mu V$)} \textit{for the signal differences} $\Delta
A$ \textit{and} $\Delta A^{\prime}$ \textit{(dashed and solid curve,
respectively). The instrumental drift has been taken into account.
It is} $\overline{\Delta A}$=\textit{2.411} \emph{$\mu$V}
\textit{($\sigma_{\Delta A}$\textit{=0.601} $\mu V$);}
$\overline{\Delta A^{\prime}}$\textit{=0.116 \emph{$\mu$V}
($\sigma_{\Delta A'}$\textit{=0.602} $\mu V$)}. \emph{$
\overline{\Delta A - \Delta A'} =
 3.81 \sigma_{\Delta
A'}$}}\label{gaussian}
\end{centering}
\end{figure}

These data are apparently at variance with electrodynamics which
expects that all of the differences of the two types should be
compatible with zero. This should be the case because, by the actual
design of the experimental box, detector A should not be affected by
the state of lighting of the source S$_{2}$. $\Delta A^{\prime }$ is
indeed compatible with the prediction of electrodynamics. The
situation is quite different for the differences $\Delta A$.
Actually, only some values (73/180) are inside the null interval,
which were discarded, while most of them (107/180) lie
outside the null interval. As it is clear from Fig.\ref{diff_no_compat_zero}%
, the differences A(S$_{1}$ on, S$_{2}$ off) - A(S$_{1}$ on, S$_{2}$
on) are positive and shifted upwards (needless to say, the stability
of power supplies was constantly checked) which means that A(S$_{1}$
on, S$_{2}$ off) $>$ A(S$_{1}$ on, S$_{2}$ on). In other words,
despite the greater number of photons in the box when both sources
are on, detector A
sees less photons than those seen when only S$_{1}$ is on\footnote{%
The detectors and the detecting circuitry were so that the higher the $\mu $%
V, the higher the number of photons collected by the detector.}. Let us
stress that it is impossible to account for this systematic effect by a
destructive interference between photons from the two sources, because the
LEDs are incoherent sources of light. We performed statistical analysis ~%
\cite{Card_Mign,ombra3} of the data and found out that there is no
compatibility between the $\Delta A^{\prime }$ set and the $\Delta A$ set
non compatible with zero as shown in Fig.\ref{gaussian}. In particular the
mean values of the two gaussian distributions are 3.81$\sigma $ apart.

\section{Interpretations}

Some not exhaustive interpretations are presented here in order to sketch
the far reaching consequences of the results of this experiment\footnote{%
For extensive interpretations you can refer to ~\cite{CASYS09,SOPO2012}}.
Although the experiment was designed so that the detector A should not be
affected by the lighting condition of the source S$_{2}$, the differences (A
(S$_{1}$ on S$_{2}$ off)-A (S$_{1}$ on S$_{2}$ on)) are convincingly
incompatible with zero. This evidence, which has been ascertained to be true
beyond any reasonable doubt, can be raised from the level of mere evidence
to the rank of physical effect if a physical cause is clearly spotted. This
experiment was explicitly designed in order to study LLI breakdown in terms
of space-time deformation and hence measures the effects of this
deformation. In this sense we can imagine that the energy of the photons
emitted by S$_{2}$ locally deforms (electrically, not gravitationally)
space-time and that this deformation expands through the aperture F$_{2}$,
reaches the photons emitted by S$_{1}$ and steers (pilots) their propagation
before they are detected by A\footnote{%
Just like the curved spacetime around the Sun curves the trajectory of the
photons from a distant star. However, while the concept that energy can
affect the geometry of space-time is the same as that in General Relativity,
this deformation has nothing to do with gravitation.}. A similar kind of
interpretation in terms of something, other than photons, moving through the
aperture F$_{2}$ can be given in terms of Bohmian Mechanics and pilot wave.
In this case, it is possible to imagine that the pilot waves of the photons
emitted by S$_{2}$ propagate through the aperture F$_{2}$ and steer the
photons emitted by the source S$_{1}$ before they reach the detector A.
According to this interpretation, this is the first experiment ever in which
a direct evidence of pilot waves is achieved. The similarity between these
two interpretations allows to put forward the intriguing hypothesis of a
possible connection between them. In particular we can say that what is
called pilot wave is nothing but a deformation of the local space-time
geometry, intimately bound to the quantum entity considered (photon) which,
in this sense, becomes a much more complex object than the quantum
mechanical picture. In particular, with regards to the photon we can say
that most of its energy is concentrated in a tiny extent (complying with
electrodynamics, relativity and Minkowski space-time) and the rest of the
energy is used to deform the space-time surrounding it (violating
electrodynamics, not complying with relativity and hence possessing real
non-local and superluminal features). This second part of the energy is
stored in the local deformation of space-time just as the Riemann curvature
of space-time in General Relativity possesses its own energy momentum
pseudo-tensor. According to the last sentence about energy stored in the
deformation, it is possible to state that pilot waves cannot be hollow waves
any longer\footnote{%
For more extensive interpretations refer to ~\cite{CASYS09,SOPO2012}}.

\section{Conclusions and remarks}

The change of the number of photons detected by A can be read as well in
terms of the modification of the photon-photon cross section due to the
deformed space-time associated to each photon. With this last picture of
modified cross section, we report that compatible results with ours were
obtained in crossed photon-beam experiments both in the microwave range ~%
\cite{ranf1,ranf2} and with a CO2 laser ~\cite{card1,card2}. Crossed laser
beams are certainly a much simpler experimental set-up, however, it goes
without saying that, despite the apparent simple and very common appearance
of the equipment, the features of LLI breakdown are so peculiar, as we have
extensively pointed out, that require subtle care in tuning all the
experimental features in order to make out the anomalous effect due to
deformed space-time. However the intersecting of laser beam is a very
suitable set-up to deepen the study of different features: spatial extension
by varying horizontally and vertically the crossing region of the beams;
timing of the deformation by applying different chopper frequencies and
sampling time procedures; investigating the effects of deformed space-time
on the frequency of photons; attempts to work as close as possible to the
single photon; drawing interesting similarities between the nonlinear and
non-local effects studied by non linear optics in nonlinear media (liquid
crystals) and non linear non local effects of space-time; visualize by CCD
the deformation of the laser beam spot which is the unmistakable evidence of
deformed space-time through which photons propagate as already tried and
reported in ~\cite{card2}.

\end{document}